\documentclass[12pt]{article}
\usepackage{amssymb,hyperref}

\newcommand{\om}{\omega}
\newcommand{\Tr}{{\rm Tr}}
\newcommand{\la}{{\lambda}}
\newcommand{\upd}{{\rm d}}
\renewcommand{\O}{O}

\begin{document}
\title{Unified analysis of terminal-time control in classical and quantum systems}
\author{Alexander Pechen and Herschel Rabitz}
\date{24 June 2010}
\maketitle
\center{\large Department of Chemistry, Princeton University, Princeton, NJ, USA}

\abstract{Many phenomena in physics, chemistry, and biology involve seeking an optimal control to maximize an objective for a classical or quantum system which is open and interacting with its environment. The complexity of finding an optimal control for maximizing an objective is strongly affected by the possible existence of sub-optimal maxima. Within a unified framework under specified conditions, control objectives for maximizing at a terminal time physical observables of open classical and quantum systems are shown to be inherently free of sub-optimal maxima. This attractive feature is of central importance for enabling the discovery of controls in a seamless fashion in a wide range of phenomena transcending the quantum and classical regimes.}

\section{Introduction}
The control of quantum phenomena is a subject garnering increasing attention, and the allied classical analog has a long history with rich applications. Controlled systems naturally appear in physics~\cite{Shulz,R1}, chemistry~\cite{chemistry} and biology~\cite{biology,biology2}, and in many cases the controlled systems are open to interact with their environments. Examples include laser driven selective atomic or molecular excitations in the presence of an external bath~\cite{REF1}, laser cooling~\cite{REF1b}, manipulation by quantum dots interacting with a reservoir of nuclear spins~\cite{REF2}, control of chemical reactions in solution~\cite{REF3}, adaptation of an organism's population to a natural or artificial environment~\cite{REF4}, etc. For control of atomic or molecular scale physical systems, quantum mechanics is a common description, whereas for biology the classical regime generally provides a suitable framework. In chemistry and other areas, either formulation may be appropriate, depending on the specific situation.

The practical contexts for seeking control typically involve either laboratory experiments or natural circumstances (e.g., an organism's population evolution driven by environmental pressure). A control problem may be expressed as maximization of an objective function $J[u]$, which is the desired property $J$ for optimization by means of the control $u$ (the problem of minimizing an objective $J$ is equivalent to maximizing $-J$). The behavior of the objective $J$ as a function of the control determines the control landscape for the corresponding problem. Various types of control problems exist, including terminal-time problems aiming to maximize an objective determined by the state of the system at some final (terminal) time~\cite{finaltime}, tracking control aiming for a particular system observable or the system state to follow a prescribed evolution pathway~\cite{trackingcontrol}, control involving conditioning on measurement results~\cite{Wiseman2010}, time-optimal control~\cite{timeoptimal}, etc. The goal is to find an optimal control $u_*$ which produces the global maximum value $J_{\rm max}$ of the objective. Commonly maximizing $J$ entails a search starting from an initial trial control, likely far from being optimal, and sampling amongst the available controls trying to identify the best solution $u_*$. If the objective $J$ has many sub-optimal maxima, then the search for an optimal control may become stuck at a particular sub-optimal maximum with an objective value lower than the desired maximum value $J_{\rm max}$. For this reason, sub-optimal maxima are referred to as traps for a control problem. Even advanced search algorithms could suffer reduced efficiency or become hopelessly lost if the control objective has a high density of sub-optimal maxima. Therefore, the presence of sub-optimal maxima can greatly influence the complexity of seeking optimal controls and the analysis for the existence or absence of traps is of central importance for optimally controlling systems described either classically or quantum mechanically.

Heretofore, control in the classical and quantum domains generally has been viewed as separate subjects, but it is natural to explore their degree of mutual control behavior. This work considers control problems for classical and quantum systems with objectives determined by the state of the system at some final time. The paper draws the important conclusion that the corresponding open classical and quantum control landscapes share the same inherent property of lacking sub-optimal extrema based on combining, in a unified framework, three facts that (a) the space of states (i.e., probability distributions for a classical system or density matrices for a quantum system) for a classical or quantum system is convex, (b) the objective functions describing common control and optimization problems are concave, and (c) concave functions do not have sub-optimal maxima over convex domains. This finding has fundamental and practical significance, as it links the ease of searching for optimal controls in the classical and quantum domains, including in systems whose various components may transcend quantum and classical behavior.

The objective in this work is expressed as a function of the final state $\rho_{\rm f}$ of the system at final time $t_{\rm f}$. The applied control $u$ directs the evolution of the system from the initial state $\rho_{\rm i}$ at time $t_{\rm i}$ to the final state $\rho_{\rm f}$ at time $t_{\rm f}$ and specifies the value of the objective $J[u]=J(\rho_{\rm f})$, which depends, through $\rho_{\rm f}$, on the control $u(t)$. The absence of sub-optimal maxima for $J[u]$ is established under common assumptions of controllability and local surjectivity for the map $\xi: u(t)\to\rho_{\rm f}$ from the space of the dynamical controls into state space~\cite{O2,Wu09}. Controllability means that the set of dynamical controls is rich enough to produce any system state, perhaps on some coarse-grained scale; thus for a given initial state $\rho_{\rm i}$ and for any final state $\rho_{\rm f}$ there exists at least one dynamical control $u$ which steers the initial state sufficiently close (as determined by the required laboratory precision) to the final state. Local surjectivity means that locally varying controls in the neighborhood of $u$ allows for moving in arbitrary directions in the state space around $\rho_{\rm f}$ on a practically sufficient coarse-grained scale. These assumptions are sufficient (but, likely can be relaxed) to draw the conclusion about absence of traps for all $J^{(1)}$ and $J^{(2)}$ objectives for a particular control application; violation of these assumptions (that can happen for example for systems with many degrees of freedom and with limited control resources) can lead to the appearance of traps in the corresponding control landscape~\cite{Wu09-2}.

The following two general classes of objective functions are considered. The first class corresponds to seeking maximization of the average value $\langle \O\rangle$ of a desired physical observable $\O$ of the system and is characterized by objectives of the form $J^{(1)}(\rho_{\rm f})=\langle \O\rangle$. Such objectives describe a wide class of control phenomena, including creation of specific atomic or molecular states, control of chemical reactions, etc. This class of objective functions will be shown to have neither sub-optimal minima nor sub-optimal maxima. The second class describes minimization of thermodynamic observables of the form $\langle \O\rangle-TS(\rho_{\rm f})$, where $S(\rho_{\rm f})$ is the entropy of the final state and $T$ is the temperature. An example is the Helmholtz free energy $\langle H\rangle-TS(\rho_{\rm f})$, where $H$ is the Hamiltonian of the system. Minimization of such observables is equivalent to maximization of objectives of the form $J^{(2)}(\rho_{\rm f})=-[\langle \O\rangle-TS(\rho_{\rm f})]$ (e.g., for minimizing Helmholtz free energy, $J_{\rm H}^{(2)}(\rho_{\rm f})=-[\langle H\rangle-TS(\rho_{\rm f})]$). If the entropy $S(\rho_{\rm f})$ is concave, which is normally the case in relevant physical, chemical and biological control applications~\cite{concaveS,abe}, then the objective function $J^{(2)}$ will also be concave, which implies the absence of sub-optimal maxima for $J^{(2)}$ control objectives. The analysis exploits the fact that the space of all states is convex for both classical and quantum open systems and that concave functions defined over convex domains~\cite{valentine1964} have no suboptimal maxima~\cite{rockafellar}. For physical systems with non-concave entropy~\cite{nonconcave,nonconcave2} a special analysis is needed beyond the scope of this work.

Control landscapes for objectives of type $J^{(1)}$ for closed quantum systems were analyzed in the dynamic~\cite{U2} and kinematic~\cite{U1} pictures corresponding to the description of the system evolution by the Schr\"odinger equation and by unitary transformations, respectively. Under the assumptions that any unitary transformation can be produced by some control and that the map from the space of controls to unitary operators has full rank, the objective function $P_{\rm i\to f}$ for maximizing population transfer was shown to have no traps~\cite{U1}. Critical points for similar objective functions in a different context were studied in~\cite{CLrefs}. The absence of traps in the space of dynamical controls was also proved~\cite{U2} under the same assumptions. For open quantum systems, the dynamic picture corresponds to the description of the evolution by various dynamical formulations, such as Markovian and non-Markovian master equations. The kinematic picture is described by completely positive trace-preserving Kraus maps in the absence of initial correlations between the system and the environment and by more general possibly non-completely positive maps if the initial state of the system and the environment is entangled~\cite{noncpmaps}. The control landscape topology of $J^{(1)}$ objectives for finite-level open quantum systems was recently treated in the kinematic picture~\cite{O1,O2} with the evolution described by Kraus maps, and trap-free behavior was established. The present paper utilizes a distinctly different formulation unifying the analysis of the topology of classical and quantum control landscapes thereby revealing their common features.

In many situations, the dynamical behavior of classical and quantum controlled or uncontrolled systems can be different in essential ways. Importantly, the analysis in this paper does not assume and does not imply the existence of common features in such dynamical behavior. Rather, regardless of the dynamical distinctions, this paper shows that upon seeking optimal controls the classical and quantum regimes share mutual characteristics. The analysis specifically considers open systems which admit a statistical description; the analysis relies on the convex structure of the set of all states for open systems and can not be directly applied to closed systems for which the allowed sets of states are non-convex. For example, consider a closed classical system initially in a definite state with a probability distribution represented by a $\delta$-function. Then the allowed probability distributions will be $\delta$-functions, and a convex sum of two different $\delta$-functions is not a $\delta$-function. Therefore, the set of all allowed states for such a closed quantum system, as well as for a classical system with any other initial state, will be non-convex. For a closed quantum system the set of allowed states consists of unitary transformations of the initial state and is also non-convex. Nevertheless, using completely different methods one can show the absence of traps for closed quantum systems in both kinematic~\cite{U1} and dynamic pictures~\cite{U2}.

\section{Classical control landscapes} The time evolution of an open classical system is described by the corresponding phase space distribution function (probability measure). Thus, classical probability theory~(see~\cite{Shulz}, Sec.~6) forms a basis to analyze open classical control landscapes. The classical probability space is defined by the triple $(\Omega,{\cal F},P)$, where $\Omega$ is a non-empty set of the system's phase space variables, ${\cal F}$ is a $\sigma$-algebra of subsets of $\Omega$, and $P:{\cal F}\to[0,1]$ is a probability measure. A sigma-algebra ${\cal F}$ over $\Omega$ is a nonempty collection of subsets of $\Omega$ that is closed under complementation and the countable union of its elements. A probability measure over $\Omega$ satisfies the following three Kolmogorov axioms~\cite{kolmogorov1950}:

A1. $P(E)\ge 0$ for any $E\in{\cal F}$.

A2. $P(\Omega)=1$.

A3. For any countable sequence of pairwise disjoint subsets $E_1,E_2,\dots\in{\cal F}$ we have
$$
P(E_1\cup E_2\cup\dots)=P(E_1)+P(E_2)+\dots
$$
As an example, for a particle moving in three-dimensional space, the phase space is $\Omega=\mathbb R^3\times\mathbb R^3$, and a point $\omega=(p,q)\in\Omega$ in this space is specified by the momentum $p$ and the position $q$ of the particle. For a more complex system $\Omega$ would encompass all relevant phase space variables.

Any state of a classical open system can be represented by a probability measure $P$ over its phase space $\Omega$, or equivalently, by the distribution function $\rho(\om)$ which is a non-negative function such that the corresponding probability measure is $P(\upd\om)=\rho(\om)\upd\om$. The set ${\cal P}_\Omega$ of all probability distributions over $\Omega$ is convex. That is, for any two probability distributions $\rho_0$ and $\rho_1$ and for any $\la\in[0,1]$, the convex sum $\rho_\la:=\la \rho_1+(1-\la)\rho_0$ is also a probability distribution in ${\cal P}_\Omega$.

The physical properties of a classical open system are prescribed by random functions, i.e., by real measurable functions defined on the phase space $\Omega$. Any such function (observable) $O:\Omega\to\mathbb R$ depends on the system's phase space variables $\om$. Physically measurable quantities for an open system are the average values of the corresponding random functions. If the classical open system at the initial time $t_{\rm i}$ is in the state characterized by the probability distribution function $\rho_{\rm i}(\om)$, then the average value of the property $O$ is
$$
\langle O\rangle_{t_{\rm i}}=\int\limits_\Omega O(\omega)\rho_{\rm i}(\om)\,\upd\om=:\langle\rho_{\rm i},O\rangle
$$
Under the action of a control $u$ the system evolves into a state at time $t_{\rm f}$ with the new probability distribution function $\rho_{\rm f}(\om)$ ($\rho_{\rm f}(\om)$ depends on $u$ and $\rho_{\rm i}$). The average value of the property at $t_f$ is referred to as the classical type one landscape $J^{(1)}_{\rm c}$
\begin{equation}\label{s:eq1}
J^{(1)}_{\rm c}(\rho_{\rm f})=\int\limits_\Omega O(\omega)\rho_{\rm f}(\om)\,\upd\om=\langle\rho_{\rm f},O\rangle\, ,
\end{equation}
and the objective is to find a suitable optimal control $u_*$ that maximizes $J^{(1)}_{\rm c}(\rho_{\rm f})$. The more general type two non-linear objective landscapes of the form $J^{(2)}_{\rm c}(\rho_{\rm f})=-[\langle \O\rangle-TS(\rho_{\rm f})]$ will be also treated by the analysis below with the goal of maximizing the objective.

\section{Quantum control landscapes} The quantum mechanical landscape description here will specifically parallel that of the classical treatment above to clearly demonstrate their common landscape topology. The state of an $n$-level quantum system is represented by positive, unit trace $n\times n$ density matrices. Denoting ${\cal M}_n=\mathbb C^{n\times n}$ as the set of all $n\times n$ complex matrices, then the set of all density matrices for an $n$-level quantum system is defined as ${\cal D}_n:=\{\hat\rho\in {\cal M}_n\,|\, \hat\rho\ge 0, \Tr\hat\rho=1\}$, where $\Tr$ denotes trace. The set of all density matrices ${\cal D}_n$ is a convex set, since for any $\la\in[0,1]$ and for any two density matrices $\hat\rho_0$ and $\hat\rho_1$, their convex combination $\hat\rho_\la=(1-\la)\hat\rho_0+\la\hat\rho_1$ is also a density matrix in ${\cal D}_n$.

The physical observables of a quantum system are represented by Hermitian operators $\hat O\in{\cal M}_n$ in the system's Hilbert space ${\cal H}$. If at the initial time $t_{\rm i}$ the system is in the state $\hat\rho_{\rm i}$, then the average value of $\hat O$ will be
$$
 \langle\hat O\rangle_{t_{\rm i}}=\Tr [\hat\rho_{\rm i}\hat O]=\langle \hat\rho_{\rm i},\hat O\rangle
$$
where $\langle \hat X,\hat Y\rangle:=\Tr[\hat X^\dagger\hat Y]$, for any $\hat X,\hat Y\in{\cal M}_n$, denotes the inner product in ${\cal M}_n$ (note that $\hat\rho_{\rm i}$ is Hermitian).

A wide range of control problems for open quantum systems can be formulated as maximization of the average value of a suitable target operator $\hat O$. The control $u(t)$ (e.g., a tailored laser field~\cite{R1} or incoherent non-equilibrium environment~\cite{R2,R3}) induces evolution of the initial density matrix into some final state $\hat\rho_{\rm f}=\hat\rho_{\rm f}(u;\hat\rho_{\rm i})$. The final state may depend linearly on $\hat\rho_{\rm i}$ if initially the system and the environment are uncorrelated and non-linearly if the initial preparation is correlated~\cite{romero}. The control landscape analysis below does not use specific properties of the dynamics and is equally suitable for both linear and non-linear classical and quantum evolutions. In the quantum case, the goal is to find an optimal control $u_*(t)$ which maximizes the expected value $\langle\hat O\rangle_{t_{\rm f}}=\Tr [\hat\rho_{\rm f}\hat O]$ at the final time $t_{\rm f}$. The corresponding quantum mechanical type one objective landscape function $J^{(1)}_{\rm q}$ is
\begin{equation}\label{Jq}
J^{(1)}_{\rm q}(\hat\rho_{\rm f})=\Tr[\hat\rho_{\rm f}\hat O]=\langle\hat \rho_{\rm f},\hat O\rangle\,.
\end{equation}
Non-linear type two objective landscapes have the form $J^{(2)}_{\rm q}(\hat\rho_{\rm f})=-[\langle \hat O\rangle-TS(\hat\rho_{\rm f})]=-[\Tr[\hat\rho_{\rm f}\hat O]-TS(\hat\rho_{\rm f})]$ where $S(\hat\rho_{\rm f})$ is the quantum mechanical entropy. The landscape analysis below is performed for situations when entropy is a concave function of $\hat\rho_{\rm f}$. The explicit form of the entropy is not important, and one can use either the von Neumann entropy $S(\hat\rho)=-\Tr[\hat\rho\log\hat\rho]$, Tsallis, or any other concave form~\cite{concaveS,abe} with the same conclusion for the topology of the control landscape. For non-concave entropy forms~\cite{nonconcave,nonconcave2} a special analysis is needed beyond the scope of this work.

\section{Unified control landscape topology} The classical and quantum open system type one objective landscapes, respectively $J^{(1)}_{\rm c}(\rho_{\rm f})=\langle \rho_{\rm f},O\rangle$ and $J^{(1)}_{\rm q}(\hat\rho_{\rm f}) =\langle \hat\rho_{\rm f},\hat O\rangle$, share a common linear dependence upon their associated states, respectively, $\rho_{\rm f}$ and $\hat\rho_{\rm f}$. The type two objective landscapes $J^{(2)}_{\rm c}$ and $J^{(2)}_{\rm q}$ also have a common structure as concave functions of their respective states. In addition, the sets of all states are convex for both open classical and quantum systems, respectively, ${\cal P}_\Omega$ and ${\cal D}_n$.

Linear functions over a convex domain do not have suboptimal minima or maxima, and all of their minima and maxima are global; moreover, each level set of such a function is a connected set (see Appendix). This result leads to the general conclusion about the topology of $J^{(1)}_{\rm c}$ and $J^{(1)}_{\rm q}$: (a) the control landscape does not have suboptimal minima or maxima and therefore is trap free and (b) each level set of the control landscape is connected including at the global minima and maxima. Similarly, the landscapes $J^{(2)}_{\rm c}$ and $J^{(2)}_{\rm q}$ are concave functions of $\rho_{\rm f}$ and $\hat\rho_{\rm f}$, respectively, and are defined on their associated convex domains ${\cal P}_\Omega$ and ${\cal D}_n$. Concave non-linear functions defined over a convex domain do not have suboptimal maxima, and only global maxima are allowed~\cite{rockafellar} (see Appendix). Thus, type two control landscapes $J^{(2)}_{\rm c}$ and $J^{(2)}_{\rm q}$ for open classical and quantum systems do not have suboptimal maxima on ${\cal P}_\Omega$ and ${\cal D}_n$, respectively, and therefore they are also trap-free.

These results establish the important basic topological properties of the $J^{(1)}$ and  $J^{(2)}$ control landscapes for completely controllable open classical and quantum systems. The essential feature entering the analysis above is the convex structure of all allowed states, which is specific for open systems. If the system is controllable but significant constraints are placed on the controls, then they can restrict the set of all available states to a non-convex set possibly producing \textit{false} traps on the nominally trap free landscape. Such false traps can be avoided by removing the constraints. If the system is uncontrollable, then the corresponding control landscape can have \textit{real} traps even when the controls are unconstrained.

The analysis of~\cite{O1,O2} was able to reveal the presence of saddles in the control landscapes for open quantum systems. Furthermore, for closed quantum systems both the kinematic and dynamic analyses also identified the presence of non-trapping saddles. In general, different formulations of the controlled evolution (e.g., for the quantum case either in terms of the final state, as a Kraus map, or in terms of dynamical controls) can reveal distinct aspects of the control landscapes.

\section{Conclusions} This work establishes, in a unified fashion, the absence of any sub-optimal trapping extrema for controlled open classical and quantum systems under suitable conditions and implies the existence of a seamless transition upon seeking control running from the nano-scale quantum regime out to classical systems at the micro-scale and beyond. The broad scope of this conclusion is in keeping with its foundation resting on (i) basic physical characteristics of quantum and classical observables and (ii) simple principles from convexity analysis. It is surprising that this important conclusion steams from drawing together these well established, yet hitherto, unconnected components. The lack of sub-optimal landscape extrema implies that any search algorithm which can distinguish up from down directions on the landscape should be able to identify the absolute best value for any $J^{(1)}$ or $J^{(2)}$ objective in a controlled system. This statement does not necessarily imply that the search will be efficient, but it will not be impeded by sub-optimal extrema and therefore the search will eventually find a globally optimal solution. Conversly, if for some open system such an algorithm gets stuck at a sub-optimal extremum, then that finding implies that the system is operating with significant constraints on the dynamical controls.

\section*{Appendix} The mathematical foundation for the conclusion in the body of the paper about trap free structure of the control landscapes for open classical and quantum systems rests on the fact that any maximum (resp., minimum) for any concave (resp., convex) function defined over a convex set~\cite{valentine1964} is a global maximum (resp., minimum)~\cite{rockafellar}. For convenience to the reader, below we give a proof of this simple but crucial result.

A function $f:X\to\mathbb R$, where $X$ is a topological space, has a {\it suboptimal} maximum (resp., minimum) at $x_0$ if there exists an open neighborhood $U(x_0)\subset X$ of $x_0$ such that $\forall x\in U(x_0):$ $f(x)\le f(x_0)$ (resp., $f(x)\ge f(x_0)$) and $x_0$ is not a global maximum, i.e., $\exists x_1\in X$ such that $f(x_1)>f(x_0)$ (resp., $x_0$ is not a global minimum, i.e., $\exists x_1\in X$ such that $f(x_1)<f(x_0)$). A suboptimal maximum (minimum) is distinguished from {\it global} maxima (minima) for the function $f$. A global maximum (resp., minimum) is defined as a point $x_0\in X$ such that $f(x)\le f(x_0)$ (resp., $f(x)\ge f(x_0)$) for all $x\in X$.

A subset $X\subset\mathbb V$ of a linear space $\mathbb V$ is a convex set if for any $x_0,x_1\in X$ and any $\lambda\in[0,1]$ the point $x_\lambda:=(1-\lambda) x_0+\lambda x_1$ is in $X$. A function $f:X\to\mathbb R$, where $X$ is a convex set, is called concave (resp., convex), if $\forall x_0,x_1\in X$ and $\forall\la\in[0,1]$: $f((1-\la)x_0+\la x_1)\ge (1-\la)f(x_0)+\la f(x_1)$ (resp., $f((1-\la)x_0+\la x_1)\le (1-\la)f(x_0)+\la f(x_1)$). Note that a linear function $f:X\to\mathbb R$ satisfies $f((1-\la)x_0+\la x_1)=(1-\la)f(x_0)+\la f(x_1)$ and therefore any linear function is convex and concave.

Let $X$ be a convex set and $f: X\to\mathbb R$ be a concave function. Then $f$ does not have suboptimal maxima on $X$ as shown below by {\it reduction ad absurdum}. The absence of suboptimal minima for a convex function can be shown in the same way.

Suppose there exists a point $x_0\in X$ which is a suboptimal maximum for $f$. By the definition of a suboptimal maximum, this means that (1) $f(x)\le f(x_0)$ in some neighborhood $U(x_0)\subset X$ of $x_0$ and (2) there exists $x_1\in X$ such that $f(x_1)>f(x_0)$ (because $x_0$ is not a global maximum). Since $X$ is a convex set, for any $\la\in[0,1]$ the point $x_\la:=(1-\la)x_0+\la x_1$ is in $X$. Since $f$ is concave, we have
\begin{eqnarray*}
 f(x_\la)&=&f((1-\la)x_0+\la x_1)\ge (1-\la)f(x_0)+\la f(x_1)\\
 && >(1-\la)f(x_0)+\la f(x_0)=f(x_0)
\end{eqnarray*}
Here the first inequality follows from the definition of a concave function; the second inequality holds since $f(x_1)>f(x_0)$. Thus, for any $0<\la\le 1$,
$$
 f(x_\la)>f(x_0)
$$
For arbitrarily small $\la$, the point $x_\la$ will be arbitrarily close to $x_0$, and therefore this inequality contradicts the assumption that $x_0$ is a suboptimal maximum. Similarly one can prove absence of suboptimal minima for a convex function. In particular, a linear function $f$ does has neither suboptimal minima nor maxima over a convex domain.

The convex structure of the domain $X$ is important in this analysis. For example, a concave function initially defined over a convex domain can have suboptimal maxima after restriction to a non-convex subset.

For a function $f:X\to\mathbb R$, its {\it level set} corresponding to the value $a\in\mathbb R$ is defined as the set $X_a:=\{x\in X\,|\,f(x)=a\}\equiv f^{-1}(a)$. With this definition, each level set of a linear function $f:X\to\mathbb R$ with a convex domain $X\subset\mathbb V$ is connected. In fact, let $x_0$ and $x_1$ be any two points on the same level set, i.e. $f(x_0)=f(x_1)$. By the definition of a convex set, for any $\lambda\in [0,1]$ the point $x_\lambda=(1-\lambda)x_0+\lambda x_1$ is in $X$. The function $f$ is linear and therefore $f(x_\lambda)=(1-\lambda)f(x_0)+\lambda f(x_1)=f(x_0)$. Thus the segment $\{x_\lambda\,|\,\lambda\in[0,1]\}$ belongs to the same level set as $x_0$ and $x_1$ and
connects the points $x_0$ and $x_1$.

\section*{Acknowledgments}The authors acknowledge support from the NSF and ARO. A. Pechen also acknowledges partial support from RFFI 08-01-00727-a and NS-7675.2010.1.

\end{document}